\documentclass{osa-article}

\journal{osajournal}
\usepackage{ucs}
\usepackage{amsmath}
\usepackage{amsfonts}
\usepackage{amssymb}
\usepackage{makeidx}
\usepackage{graphicx}
\usepackage{xcolor} % to change the color of the text.
\usepackage{ulem} % to use sout
\usepackage{bm} % bold mathcal.
\usepackage{dsfont}
\usepackage{tipa} % for \textcrb
\usepackage{array}
\usepackage{bigints}
\usepackage{enumitem} % for "description" lists
\usepackage{environ} % to have "\Body" available for my own environments
\usepackage{longtable}

\allowdisplaybreaks % to allow equation split from one page to the next.

	\renewcommand{\emph}[1]{\textit{#1}}
	\newcommand{\eg}{\textit{e.g.} }
	\newcommand{\cf}{\textit{cf.} }
	
	\newenvironment{soe}
				{\left\lbrace\begin{array}{@{}l@{}}}
				{\end{array}\right.}
	\NewEnviron{subs}[1]
				{
				\begin{subequations}
				\label{#1}
				\begin{align}
				\BODY
				\end{align}
				\end{subequations}
				}
	\newcommand{\vect}[1]{\boldsymbol{\mathbf{#1}}} 
	
	\newcommand{\dir}[1]{\hat{\boldsymbol{\mathbf{#1}}}} 
	\newcommand{\mat}[1]{{\underline{\underline{#1}}\,}}
	
	\newcommand{\rpos}{\mathbf{r}}
		\newcommand{\rposp}{\mathbf{r}'}

\def\XXint#1#2#3{{\setbox0=\hbox{$#1{#2#3}{\int}$}
     \vcenter{\hbox{$#2#3$}}\kern-.5\wd0}}

\newcommand{\xdir}{$\dir{x}$ }
\newcommand{\ydir}{$\dir{y}$ }
\newcommand{\zdir}{$\dir{z}$ }

	\newcommand{\Pabs}{P_\text{abs}}
	\newcommand{\Pabsq}{P_{\text{abs},q}}
\newcommand{\GF}{\underline{\underline{\Gamma}}}
	\newcommand{\GFej}{{\underline{\underline{\Gamma}}^{EJ}}}
	\newcommand{\GFem}{{\underline{\underline{\Gamma}}^{EM}}}
	\newcommand{\GFhj}{{\underline{\underline{\Gamma}}^{HJ}}}

\newcommand{\rone}{\rpos_1}
\newcommand{\rtwo}{\rpos_2}
\newcommand{\rthree}{\rpos_3}
\newcommand{\rfour}{\rpos_4}

\newcommand{\fieldcorr}{\mat{C}_\text{em}(\rone, \rtwo)}

\newcommand{\fieldcorrnor}{\mat{C}_\text{em}}

	\newcommand{\fieldcorrEE}{\mat{C}_\text{em}^{EE}(\rone, \rtwo)}
	\newcommand{\fieldcorrEH}{\mat{C}_\text{em}^{EH}(\rone, \rtwo)}
	\newcommand{\fieldcorrHE}{\mat{C}_\text{em}^{HE}(\rone, \rtwo)}
	\newcommand{\fieldcorrHH}{\mat{C}_\text{em}^{HH}(\rone, \rtwo)}
	
		\newcommand{\fieldcorrqEE}{\mat{C}_{\text{em},q}^{EE}(\rone, \rtwo)}
		\newcommand{\fieldcorrqEH}{\mat{C}_{\text{em},q}^{EH}(\rone, \rtwo)}
		\newcommand{\fieldcorrqHE}{\mat{C}_{\text{em},q}^{HE}(\rone, \rtwo)}
		\newcommand{\fieldcorrqHH}{\mat{C}_{\text{em},q}^{HH}(\rone, \rtwo)}

\newcommand{\Cabsq}{\mat{C}_{\text{abs},q}(\rone, \rtwo)}
\newcommand{\Cabsnor}{\mat{C}_\text{abs}}
\newcommand{\Cabsqnor}{\mat{C}_{\text{abs},q}}

	\newcommand{\Cabsqjm}{\mat{C}_{\text{abs},q}^{JM}(\rone, \rtwo)}

		\newcommand{\Cabsqjjconj}{\big(\mat{C}_{\text{abs},q}^{JJ}\big)^*(\rone, \rtwo)}
		\newcommand{\Cabsqjmconj}{\big(\mat{C}_{\text{abs},q}^{JM}\big)^*(\rone, \rtwo)}
		\newcommand{\Cabsqmjconj}{\big(\mat{C}_{\text{abs},q}^{MJ}\big)^*(\rone, \rtwo)}
		\newcommand{\Cabsqmmconj}{\big(\mat{C}_{\text{abs},q}^{MM}\big)^*(\rone, \rtwo)}

\renewcommand{\matrix}[1]{\left[\begin{array}{@{}l@{}} #1 \end{array}\right]}

\renewcommand{\iiint}{\int \hspace{-0.25cm} \int \hspace{-0.25cm} \int}

\newcommand{\Imat}{\mat{1}}

\newcommand{\vectcurr}{\matrix{\vect{J} \\ \vect{M}}}
\newcommand{\vectcurrone}{\matrix{\vect{J}_1 \\ \vect{M}_1}}
\newcommand{\vectcurrtwo}{\matrix{\vect{J}_2 \\ \vect{M}_2}}

\begin{document}

\title{General relation between spatial coherence and absorption}

\author{D. Tihon,\authormark{1,*} S. Withington,\authormark{1} E. Bailly, \authormark{2}, B. Vest, \authormark{2} and J.-J. Greffet\authormark{2}}

\address{\authormark{1}Cavendish Laboratory, University of Cambridge, Cambridge, UK.\\
\authormark{2}Universit\'{e} Paris-Saclay, Institut d'Optique Graduate School, CNRS, Laboratoire Charles Fabry,  F-91127 Palaiseau, France.}

\email{\authormark{*}dt501@cam.ac.uk} %% email address is required



\begin{abstract*}
Despite the fact that incandescent sources are usually spatially incoherent, it has been known for some time that a proper design of a thermal source can modify its spatial coherence. A natural question is whether it is possible to extend this analysis to electroluminescence and photoluminescence. A theoretical framework is needed to explore these properties. In this paper, we extend a general coherence-absorption relation valid at equilibrium to two non-equilibrium cases: luminescent bodies and anisothermal bodies. We then use this relation to analyse the differences between the isothermal and anisothermal cases {\color{black} and to study the near-field emission of an electroluminescent source}.
{\color{blue}\copyright 2020 Optica Publishing Group. Users may use, reuse, and build upon the article, or use the article for text or data mining, so long as such uses are for non-commercial purposes and appropriate attribution is maintained. All other rights are reserved. The published version of the paper is available at https://doi.org/10.1364/OE.405484.}
\end{abstract*}

%%%%%%%%%%%%%%%%%%%%%%%%%%  body  %%%%%%%%%%%%%%%%%%%%%%%%%%
\section{Introduction}
Light emission due to spontaneous emission by a thermalized ensemble of emitters is a priori expected to be spatially incoherent. Given that spontaneous emission takes place with a random phase, the fields emitted by different emitters are not expected to interfere. However, it is well-known that spatial coherence can emerge upon propagation in vacuum as described by Zernike-van Cittert theorem~\cite{Born_1991}. Spatial coherence across the surface of a planar thermal emitter has also been reported some time ago in~\cite{Carminati_1999,Greffet_2002,Lau_2007}, where the correlation was attributed to the excitation of surface waves. 

More recently, interferences produced by two slits illuminated by photoluminescent molecules on a silver film optically pumped has been reported~\cite{Aberra_2012}. The origin of the spatial coherence is still debated. It has been suggested~\cite{Aberra_2012} that the strong coupling between the molecules and surface plasmons propagating along a metallic thin film plays a role. New experiments with a different system points to the filtering effect produced by the delocalized modes \cite{Shi_2014, Shi_2014_b}. Previous experiments had shown how to take advantage of surface plasmons or guided modes to tailor the source directivity~\cite{Kock_1990,Erdogan_1991,Greffet_2002}. All these experiments showing interferences or directivity reveal the presence of spatial coherence which can be fully characterized by the cross-spectral density tensor $\fieldcorrnor^{EE}(\rone, \rtwo, \omega)$ \cite{Born_1991, Tervo_2004}. Nevertheless, a general framework to study this quantity in non-equilibrium situations corresponding to typical light sources is still lacking. 

The understanding of spatial coherence is more advanced in the field of incandescent emission. The cross-spectral density can be directly computed using the fluctuation-dissipation theorem to model the fluctuating currents~\cite{Twiss_1955,Rytov_1989,Carminati_1999}. A powerful alternative formulation has been introduced by Rytov for systems in equilibrium. He considered the absorption by a body illuminated by two deterministic phase-locked monochromatic sources located at two points $\rone, \rtwo$ which generate an interference pattern in the absorbing body. To account for the effect of interferences on absorption, Rytov introduced a complex tensor quantity, the \emph{mixed losses} ~\cite{Rytov_1989, Saklatvala_2007, Craeye_2014, Tihon_2016}, which can be measured experimentally~\cite{Thomas_2011, Moinard_2017}. Rytov showed that, for incandescent emitters at uniform temperature, the cross-spectral density tensor of the fields is directly proportional to the complex conjugate of the mixed losses~\cite{Rytov_1989}. This result has been discussed independently in the context of microwave radiometry~\cite{Mamouni_1991,Craeye_2005, Warnick_2008}. It allows measuring absorption to infer spatial coherence of the fields thermally emitted in thermodynamic equilibrium. This type of measurements has been developped in recent years and it has been analysed in the framework of a modal description of the absorption process~\cite{Thomas_2011,Moinard_2017,Withington_2017}.

In this paper, we extend the coherence-absorption relation to a medium which is not in thermodynamic equilibrium. The first extension of the coherence-absorption relation deals with electroluminescence and photoluminescence. We introduce a natural framework for future studies of emission and absorption by molecules and semiconductors. The second extension deals with the spatial structure of the spatial coherence. Here, we compare the spatial coherence of the fields emitted by hot bodies for two cases:  thermodynamic equilibrium and anisothermal case. The difference between both is illustrated with a hot plasmonic surface when the background is at thermal equilibrium or not. In this paper, we use a spectral approach and assume an $\exp(-j\omega t)$ time dependency of the fields and currents. 

\section{Spatial correlation of luminescent fields}
Let us first derive the correlation of thermal fields emitted by a structure in the framework of fluctuational electrodynamics~\cite{Rytov_1989}. First, we use the fluctuation-dissipation theorem to derive the fluctuating currents from the knowledge of the losses. Then, knowing the sources, the fields and their correlations can be computed. The ability of fluctuational electrodynamics to model light emission by thermalized ensembles of emitters such as hot electrons in metals, pumped molecules or semiconductors can be found in~\cite{Leo_2019}.

Consider a structure described through its nonlocal position-dependent complex permittivity $\varepsilon(\rpos,\rpos')$ and real permeability $\mu(\rpos,\rpos')$. The positive imaginary part $\varepsilon"(\rpos,\rpos')$ of the permittivity accounts for electric losses in the material. These losses originate from one or many independent inelastic coupling mechanisms between the electromagnetic fields and the materials, which we will refer to as \emph{coupling mechanisms}. Since these coupling mechanisms are independent, so will be their respective contributions to the total losses: $\varepsilon"(\rpos,\rpos') = \sum_q \varepsilon_q"(\rpos,\rpos')$~\cite{Greffet_Apr2018}. For example, in a direct bandgap semiconductor, two main coupling mechanisms can be highlighted. The first one, $\varepsilon_l$, corresponds to the electron-phonon interaction. The second one, $\varepsilon_\textit{ib}$, corresponds to interband excitation of an electron, which can be seen as the generation of an electron-hole pair. Similarly, visible absorption in a metal can be attributed to intraband absorption often described by a Drude model and interband absorption due to the d-bands and accounted for by a different contribution.

Due to the interactions between particles leading to a thermalization on a sub ps time scale in condensed matter, a temperature can be defined to describe the particles energy distribution. Under some circumstances, different particles may have different temperatures and also different chemical potentials. For instance, a semiconductor that absorbs photons from a short light pulse has an increased number of electrons in the conduction band and holes in the valence band. As the electrons and holes thermalize in a sub ps time scale but recombine radiatively on a ns time scale, it is possible to define transient chemical potentials known as quasi-Fermi levels $\mu_c$ and $\mu_v$ respectively leading to a non-vanishing photon chemical potential defined as $\mu_\varphi=\mu_c-\mu_v$~\cite{Wurfel_1982,Greffet_Apr2018}. Note that when applying a voltage $V$ to an ideal pn junction, the photon chemical potential is given by $\mu_\varphi=eV$, with $e$ the magnitude of an electron's charge. 

Similarly, the absorption of a 100 fs visible pulse by a metallic nanoparticle results in a number of so-called hot electrons with energy much larger than the Fermi energy. These electrons thermalize by electron-electron scattering on a time scale on the order of 500 fs, forming a population with high temperatures well above the lattice temperature~\cite{Voisin_2001}. They then cool down by interaction with lattice phonons on a few ps time scale. The energy transfer between electrons and lattice phonons is described using the so-called two-temperature model~\cite{Fann_1992}. These two examples show that it is possible to define local and transient thermodynamic equilibrium for classes of particles (electrons, holes, phonons). This is possible on time scales and length scales larger than the collision time scales and length scales. 
{\color{black} It should be noted that a similar two-temperature model has been used in \cite{Weng_2018}, where the energy of the electrons is increased using a strong DC electric field. However, in the latter case, it is unclear whether local thermodynamic equilibrium is reached and thus if an equivalent temperature may provide a quantitatively good approximation of the complex electronic distribution.}

An important application concerns electroluminescence in semiconductors. According to the fluctuation-dissipation theorem~\cite{Landau_1969}, fluctuations of the current density are associated with the coupling mechanisms. The local form of the fluctuation-dissipation has been derived quantum-mechanically accounting for quasi-Fermi levels~\cite{Henry_1996}. A direct calculation of the ratio between local absorption and emission rates has been given for direct~\cite{Wurfel_1982} and indirect~\cite{Wurfel_1995} band gap semiconductors and experimentally tested for both cases \cite{Feuerbacher_1990, Schick_1992, Ferraioli_2004}. The fluctuations being uncorrelated, it leads to
\begin{equation}
\label{eqbis:02}
\begin{split}
\Big\langle
\vect{J}(\rthree) \cdot &\vect{J}^\dagger(\rfour) 
\Big\rangle =
2 \omega \sum_q \varepsilon_q"(\rthree,\rfour)\, \Theta\big(T_q(\rthree), \mu_q(\rthree)\big)  \Imat,
\end{split}
\end{equation}
with $\langle a \rangle$ the ensemble average of $a$, $\Imat$ the 3x3 identity matrix, $T_q(\rpos)$ and $\mu_q(\rpos)$ the temperature and photon chemical potential associated to the excitation of coupling mechanism $q$ in position $\rpos$ and $\Theta(T_q, \mu_q) = \hbar\omega/\{\exp[(\hbar \omega-\mu_q)/k_B T_q] -1\}$. Hereafter, we consider a local medium so that $\varepsilon_q"(\rthree,\rfour)=\varepsilon_q(\rthree)\delta(\rthree-\rfour)$ where $\delta$ is the Dirac generalized function. $\vect{A}^\dagger$ is the transpose conjugate of $\vect{A}$. Note that, by default, we use column vectors and the $\cdot$ operation is defined as the classical matrix multiplication operator, so that $\vect{A} \cdot \vect{B}^\dagger$ correspond to a rank-1 matrix while $\vect{A}^\dagger \cdot \vect{B}$ corresponds to a scalar quantity. We note that the fluctuation-dissipation relation remains valid for indirect transitions involving different types of carriers provided that these carriers have the same temperature~\cite{Wurfel_1995}. 

{\color{black}It can be noticed that zero-point fluctuations have been omitted in \eqref{eqbis:02} since we are interested in the fields emitted due to the excitation of a given coupling mechanism. Zero-point fluctuations are independent from the excitation and cannot lead to any net energy flux. In the macroscopic approach we are considering, it is not relevant to include them. A detailed discussion about cases where such fluctuations should be accounted for can be found in~\cite{Joulain_2005}. If required, the effect of zero-point fluctuations can be easily included in the following calculations by adding a $\hbar \omega/2$ term in the definition of the $\Theta$ function.} 

 We now turn to the study of the field spatial correlations. They can be described by a second order correlation function $\fieldcorrnor(\rone, \rtwo, t_1, t_2)$ that provides the correlation of the fields at different positions and different times. We consider that the temperature {\color{black}and photon chemical potential are} constant with time, so that thermal fluctuations can be considered as a time-stationary random process, leading to $\fieldcorrnor(\rone, \rtwo, t_1, t_2) = \fieldcorrnor(\rone, \rtwo, t_1-t_2, 0)$. Treating independently the different spectral components of the fields, the electric field correlation can be described using the cross-spectral density tensor \cite{Born_1991, Tervo_2004}
\begin{equation}
\label{eqbis:01}
\fieldcorrnor^{EE}(\rone, \rtwo, \omega) = 
\left\langle 
	\vect{E}(\rone, \omega) 
	\cdot 
	\vect{E}^\dagger(\rtwo, \omega)
	\right\rangle, 
\end{equation}
with  $\vect{E}(\rpos, \omega)$ the electric field at position $\rpos$ and at angular frequency $\omega$. Hereafter, the frequency dependence of the quantities used will be omitted for brevity. {\color{black} We note that this definition differs from the usual definition that can be found in, e.g. \cite{Carminati_1999, Greffet_Apr2018}. This is due to the difference in the conventions used to exploit stationariness and reduce the dimensionality of $\fieldcorrnor^{EE}(\rone, \rtwo, t_1, t_2)$ to $\fieldcorrnor^{EE}(\rone, \rtwo, \omega)$.}

The average of the fields depends on the average of the current density due to fluctuations, $\vect{J}$ which can be obtained from the fluctuation-dissipation theorem for each coupling mechanism~\cite{Landau_1969,Greffet_Apr2018}. To relate the fields of \eqref{eqbis:01} to the current density of \eqref{eqbis:02}, we use the Green's tensor $\GFej(\rone, \rtwo)$, which provides the electric field generated at $\rone$ by unit source currents located at $\rtwo$ ($\vect{E}(\rone) = \iiint \GFej(\rone, \rtwo) \cdot \vect{J}(\rtwo) dV(\rtwo)$), which straightforwardly leads to (\cf Appendix A.1)
\begin{align}
\fieldcorrEE & = \sum_q \mat{C}_{\text{em},q}^{EE}(\rone, \rtwo) \\
& = \sum_q
2 \omega  \iiint_\Omega \Theta\big(T_q(\rthree), \mu_q(\rthree)\big) \, \varepsilon_q"(\rthree)
\label{eqbis:03}
 \,
\GFej(\rone, \rthree) \cdot \big(\GFej\big)^\dagger(\rtwo, \rthree)
dV(\rthree).
\end{align}

\section{Mixed losses and relation to spatial coherence}
Now, we consider the dual problem: the same structure is illuminated by deterministic monochromatic external sources. These sources correspond to phase-locked point-like electric currents $\vect{J}_1$ and $\vect{J}_2$ located at positions $\rone$ and $\rtwo$, respectively. Since only electric losses are considered, the power density dissipated at $\rpos$ is $P(\rpos) = \omega \varepsilon"(\rpos) \vect{E}^\dagger(\rpos) \cdot \vect{E}(\rpos)/2$. Using the Green's tensor to relate the local value of the fields to the sources $\vect{J}_1$ and $\vect{J}_2$, integrating the losses over the whole structure and isolating the contribution of different excitations, the absorbed power can be cast in the form:
\begin{equation}
\label{eqbis:06}
\Pabs = 
\sum_{i,j = 1}^{2}
\vect{J}_i^\dagger
\cdot 
\Big[ \sum_q \mat{C}_{\text{abs},q}^{JJ}(\rpos_i, \rpos_j)
\Big]
\cdot
\vect{J}_j,
\end{equation} 
with
\begin{equation}
\label{eqbis:04}
\begin{split}
\mat{C}_{\text{abs},q}^{JJ}(\rpos_i, \rpos_j)
= &
\dfrac{\omega}{2} 
\iiint_\Omega  \varepsilon_q"(\rthree)
\big(\GF^{EJ}\big)^\dagger(\rthree, \rpos_i) \cdot \GF^{EJ}(\rthree, \rpos_j) 
dV(\rthree).
\end{split}
\end{equation}
The $\mat{C}_{\text{abs},q}^{JJ}$ tensor is the mixed losses tensor associated to coupling mechanism $q$.
The terms $i=j$ correspond to the self-contribution of each source to the power absorbed, while the terms $i\neq j$ accounts for the interferences between the sources. The details of the derivation of \eqref{eqbis:06} are available in Appendix A.2.

The mixed losses $\Cabsnor^{JJ} = \sum_q \mat{C}_{\text{abs},q}^{JJ}$ (or assimilated quantities) can and have already been measured experimentally at radio ~\cite{Thomas_2011} and infrared \cite{Moinard_2017} frequencies. First, $\Cabsnor^{JJ}(\rpos, \rpos)$ is measured by illuminating the structure with a single source. Then, the structure is illuminated by two phase-locked sources. By measuring the power absorbed for different relative phases between the sources and removing the self-contribution, one can find the amplitude and phase of the mixed losses ~\cite{Moinard_2017}. 

We now compare \eqref{eqbis:03} and \eqref{eqbis:04}. We consider the contribution to the fields of a given coupling mechanism $q$ taking place within a volume where $T_q$ and $\mu_q$ are uniform. For such an emissive volume, the correlation of the fields emitted is related to the mixed losses through
\begin{equation}
\label{eqbis:08}
\mat{C}_{\text{em},q}^{EE}(\rone, \rtwo) = 4 \Theta(T_q, \mu_q) 
\Big(\mat{C}_{\text{abs},q}^{JJ}\Big)^*(\rone, \rtwo)
\end{equation}
with $\mat{A}^*$ the element-by-element complex conjugate of $\mat{A}$. This result, whose derivation is available in Appendix A.3, is the central result of the letter. Indeed, under the hypothesis of local thermodynamic quasi equilibrium, Eq.~\eqref{eqbis:08} provides a direct and intuitive relation between the absorption of incident fields and the correlation of the fields emitted. It provides a technique to explore spatial coherence of many light sources. 

We note that this result can be extended to magnetic fields using magnetic sources (cf. ~\cite{Rytov_1989, Rodriguez_Aug2013} and Appendix A.3). We also note that if the different particles are at thermodynamic equilibrium (identical temperatures and vanishing photon chemical potentials), considering that $\sum_q \varepsilon_q" = \varepsilon"$, Eq.~\eqref{eqbis:03} corresponds to the usual result for incandescent fields {\color{black}first derived by Rytov \textit{et al.}}~\cite{Rytov_1989}{\color{black}, which is itself a generalization of the original Kirchhoff's law.}

\section{Numerical example}
\begin{figure}
\center
\includegraphics[width = 9cm]{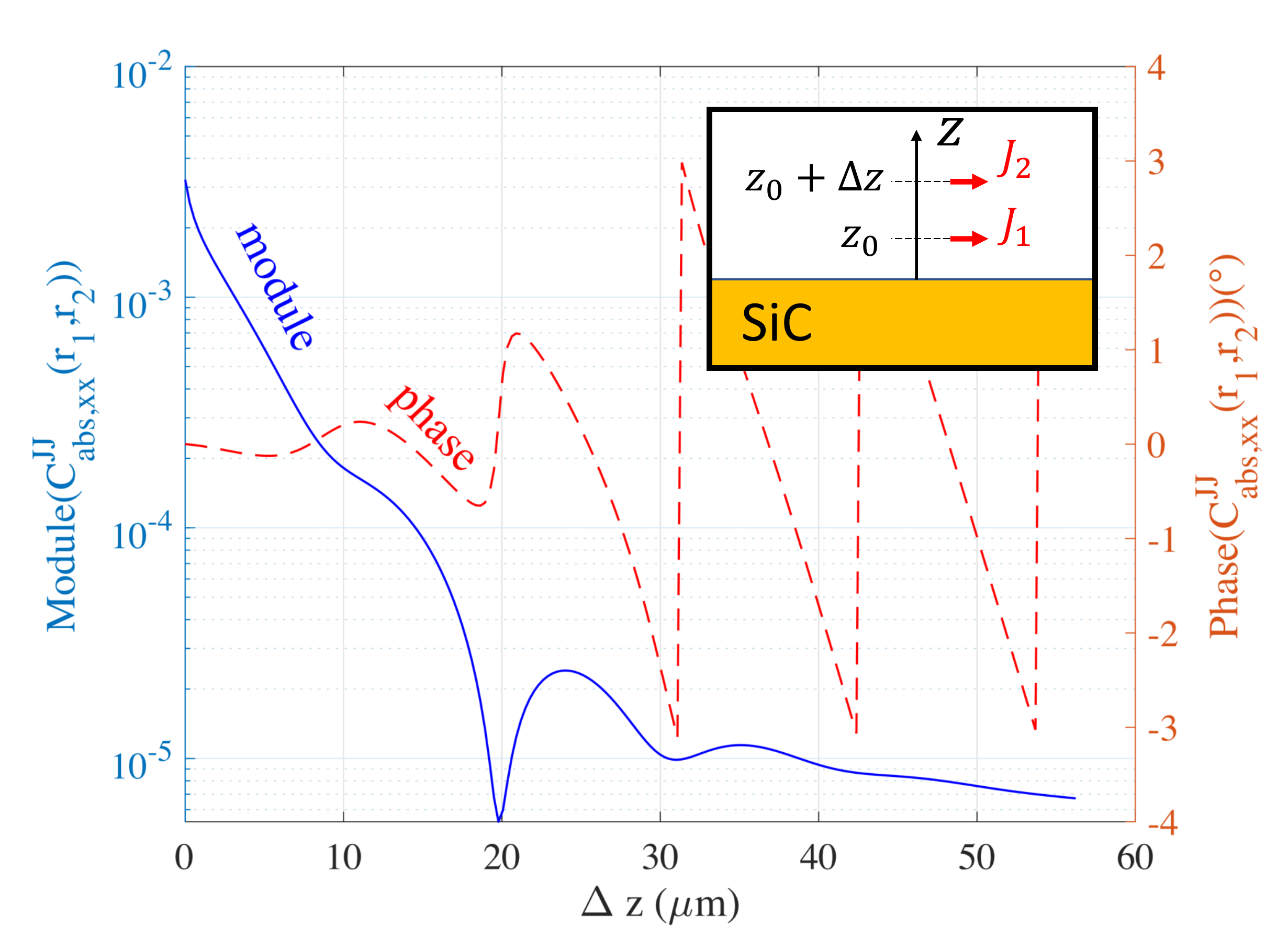}
\caption{Evolution of the mixed losses $\dir{x} \cdot \Cabsnor^{JJ}(z_0 + \Delta z \dir{z}, z_0 \dir{z}) \cdot \dir{x}$ for a SiC half-space as a function of $\Delta z$, with $z_0=0.05 \lambda$.}
\label{fig:03}
\end{figure}

We now explore the spatial structure of the field correlation for a thermal planar source at a higher temperature than its environment. At thermodynamic equilibrium, it is known that the cross spectral density tensor is proportional to the real part of the Green's tensor $\text{Re}\{\GFej(\rone, \rtwo)\}$~\cite{Rytov_1989}. We note that the concept of cross density of states (CDOS), also proportional to $\text{Re}\{\GFej(\rone, \rtwo)\}$ has been introduced recently to characterize the spatial structure of the electromagnetic modes of a given system~\cite{Caze_2013}. When dealing with a particular source, be it a thermal source or a luminescent source, not all states are excited by the source as opposed to the thermodynamic case. Hence, the cross spectral density of the emitted fields differs from the CDOS {\color{black}\cite{Eckhardt_1982, Eckhardt_1984}}. In particular, it is complex. To illustrate these differences, we consider a simple structure consisting of a silicon carbide (SiC) half-space{\color{black}. This geometry has already been extensively studied in the literature (see e.g. \cite{Carminati_1999, Lau_2007, Dorofeyev_2011}) and is thus a good platform to illustrate the relation between the mixed losses and the cross-spectral field density tensor.} At a free-space wavelength of $\lambda = 11.36$ $\mu$m, the relative permittivity of SiC is $\varepsilon_r = -7.51+0.41j$ \cite{Spitzer_1959}, so that the skin depth is about 650 nm. The SiC is illuminated by two parallel horizontal current sources located at distances of $z_0$ and $z_0 + \Delta z$ from the surface. The geometry is depicted in the inset of Figure \ref{fig:03}. If $\Delta z = \lambda/2$, one would expect a low power absorption if the sources are in phase and a high absorption if they have opposite phases, due to destructive and constructive interferences respectively.

This simple example can be extended to a continuum of positions for both sources. Associated to each pair of positions is a phase difference between the sources for which the absorbed power is maximum, and the associated $\pi$-shifted phase difference for which it will be minimum. To illustrate this, we computed the mixed losses of the sources at different heights ($\dir{x}\dir{x}$ component of the $(i,j) = (2,1)$ term of \eqref{eqbis:06} as a function of $\Delta z$ for $z_0 = 0.05\lambda$). The results can be seen in Fig. \ref{fig:03}. The magnitude of the function describes the strength of the interference-related effect while its phase corresponds to the phase difference between the two sources for which the absorbed power is maximum. 

When both sources are close to the surface, the interference-related effects are strong due to the near-field components of the dipolar field.  When $\Delta z>\lambda$, the mixed losses are mainly due to propagating waves. The resulting interferences are much weaker and the relative phase leading to maximum absorption evolves linearly with $\Delta z$.

Using the reciprocity of \eqref{eqbis:08}, we know that the correlation of the fields thermally emitted will follow the same curve, except for a constant factor and an opposite phase. At small distances from the surface, the evanescent part is the dominant contribution to the fields and the phase is close to zero. When  $\Delta z>\lambda$, the thermal fields reaching point $\rtwo$ are made of propagating waves. These propagating waves first transit through point $\rone$, where they have the same amplitude but different phase, hence the slower variation of the magnitude of the correlation function and the linear phase term that appears.

At thermodynamic equilibrium, the field correlation tensor is given by the real part of the Green tensor so that it is real (see Appendix B). \begin{equation}
\label{eqbis:07}
\fieldcorrEE = 2\Theta(T_0) \text{Re}\{\GFej(\rone, \rtwo)\}.
\end{equation}
Formula \eqref{eqbis:07} is similar to formula (1) of~\cite{Caze_2013} that defines the CDOS. It only differs in a constant factor that is due to (i) the linear proportionality factor that exists between the CDOS and the cross-spectral density function and (ii) the conventions we used to go from time to frequency domain~\cite{Greffet_Apr2018}. To understand the origin of the difference, we note that, in thermodynamic equilibrium,  the whole space is at a uniform temperature $T_0$ and black-body radiation of temperature $T_0$ is incident from infinity in order to compensate for radiation losses{\color{black}, as explained in \cite{Eckhardt_1982, Eckhardt_1984}}. In the non-equilibrium case, we considered only radiation emitted by the SiC and propagating away from it.

\begin{figure}
\center
\hspace{-0.5cm}
\begin{tabular}{cc}
\includegraphics[width = 6.5cm]{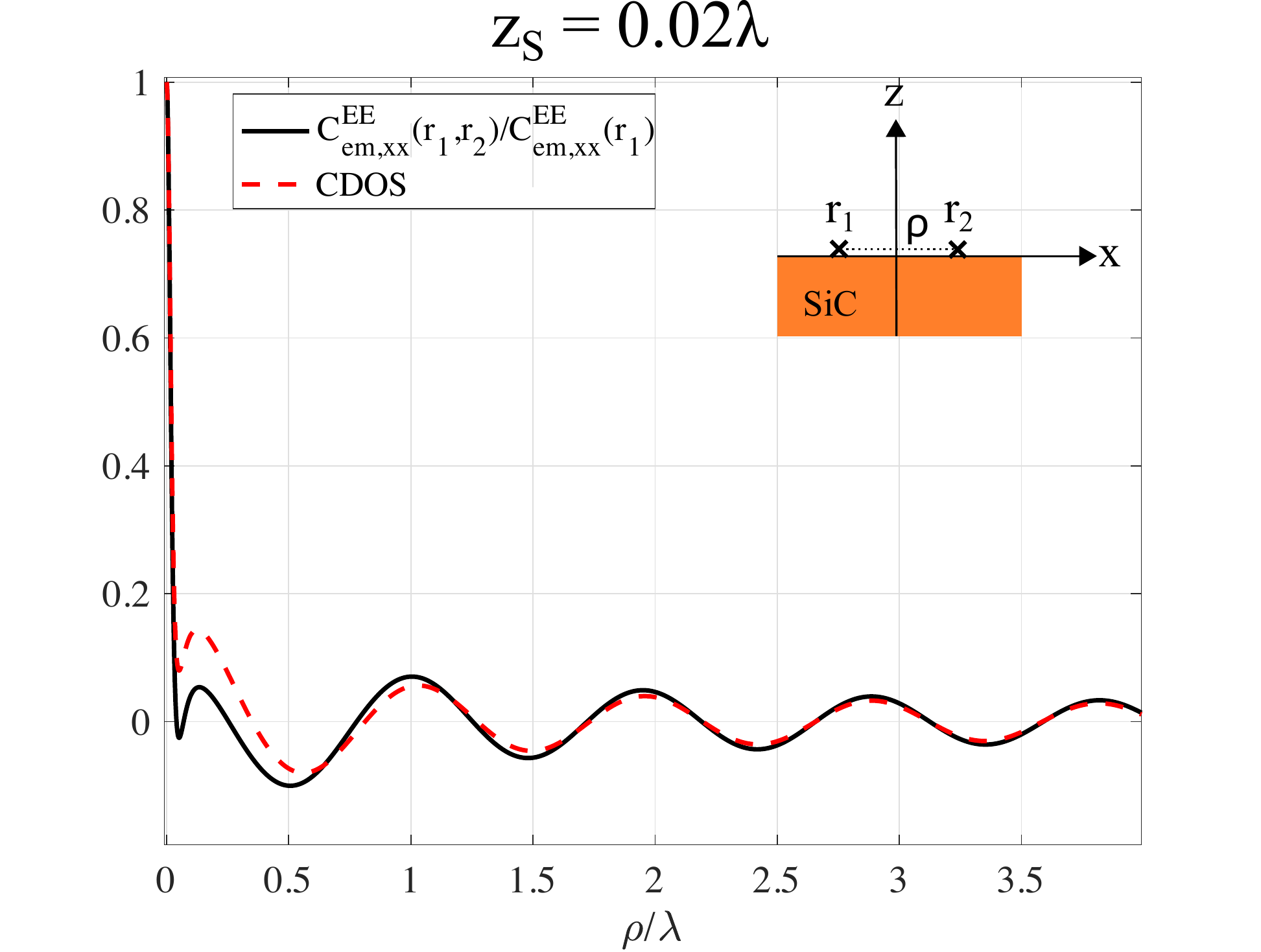} 
&
\hspace{-0.75cm}
\includegraphics[width = 5.5cm]{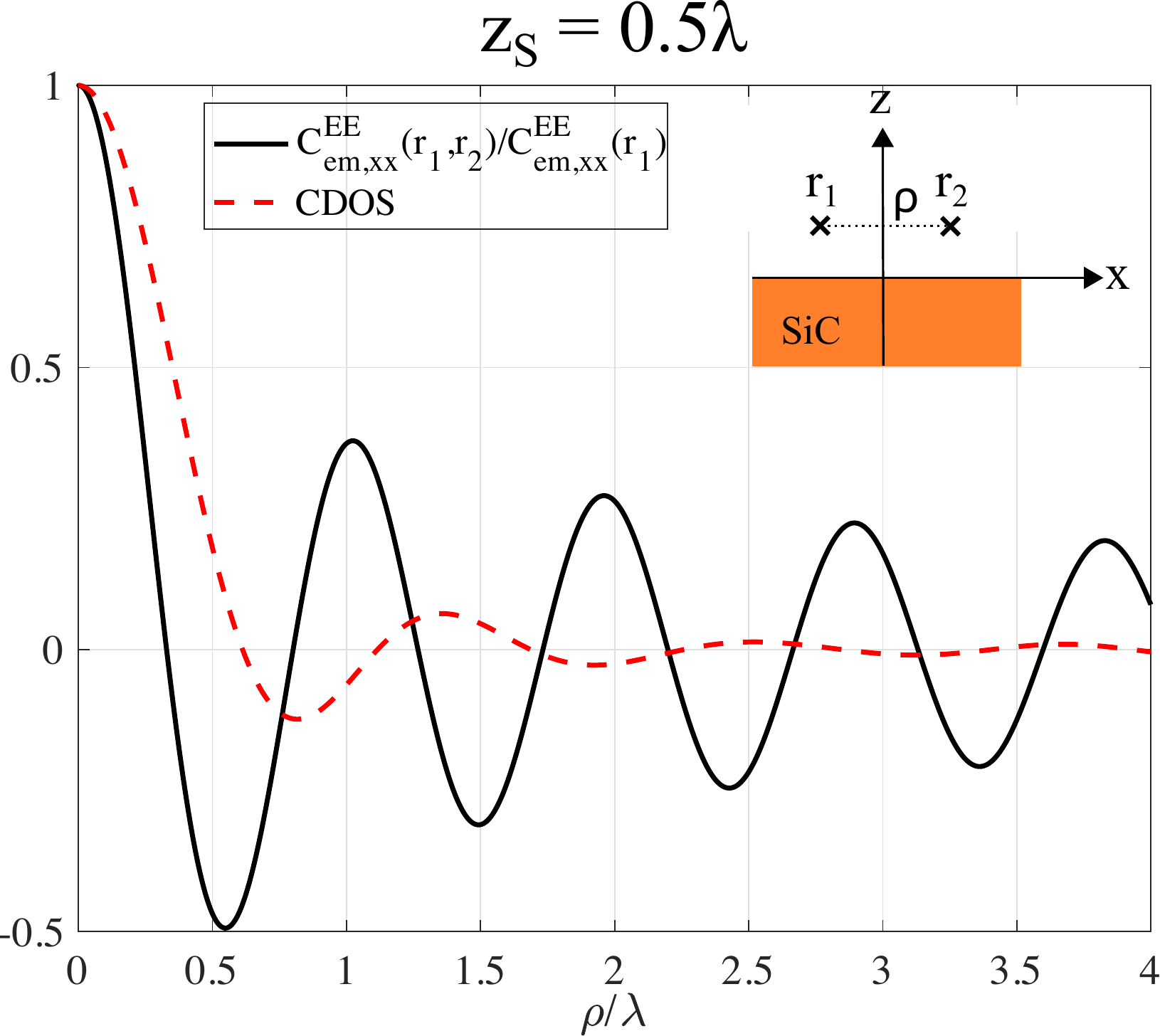}\\
(a)
&
\hspace{-0.75cm}
(b)
\end{tabular}
\includegraphics[width = 5.45cm]{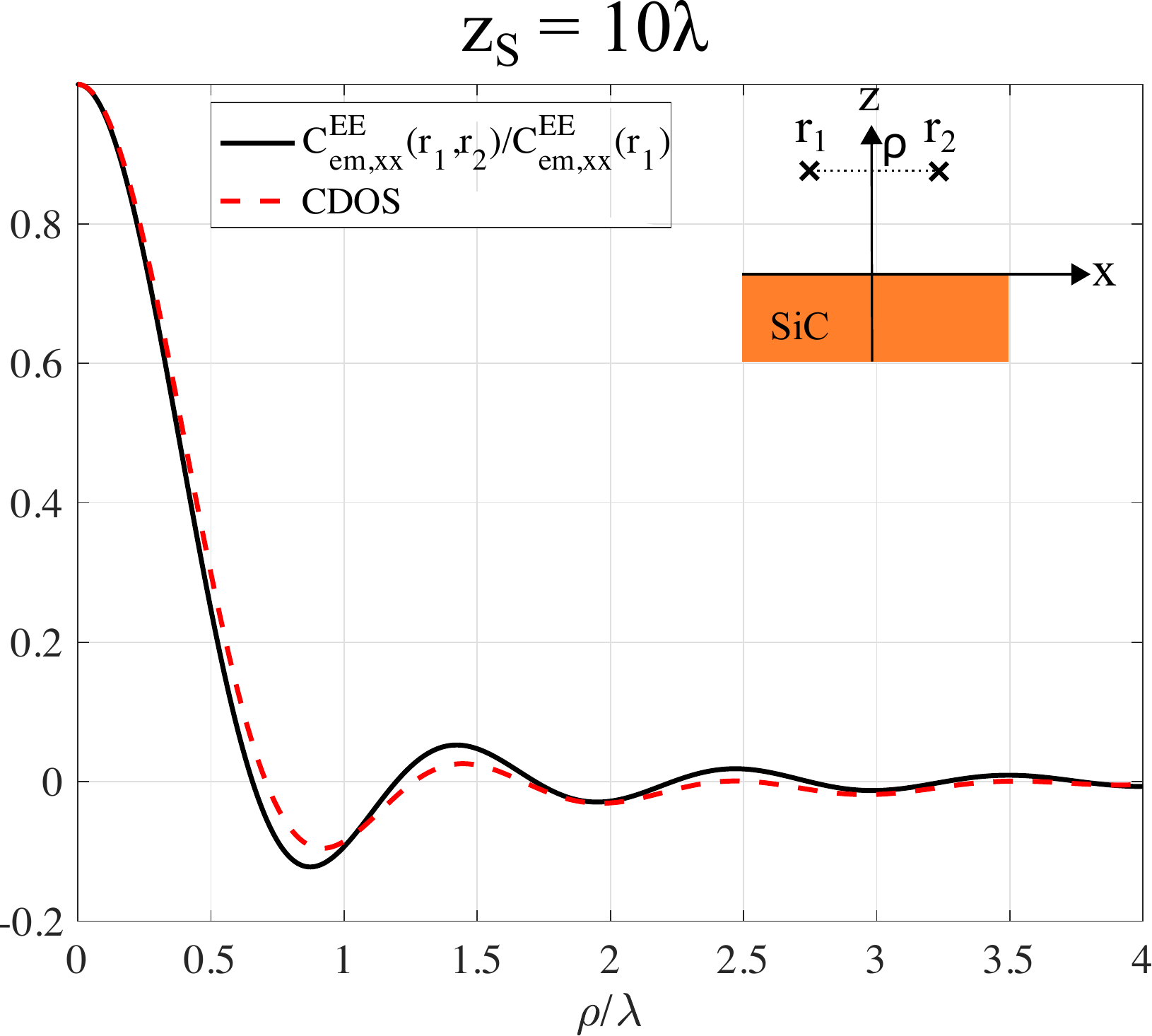}\\
(c)
\caption{Comparison of the normalized two-points correlation of the fields thermally emitted by a SiC half-space (black solid line) with the normalized CDOS (red dashed line) as a function of the lateral distance $\rho$ between the points (a) in the close vicinity, (b) in a intermediate distance and (c) at a large distance from the surface.}
\label{fig:01}
\end{figure}

To illustrate the difference between a cold space and a space at thermodynamic equilibrium, we consider the correlation of the $\dir{x}$-directed thermal fields generated by SiC in a cold space (zero temperature) and compare it with the CDOS for varying distances $z_s$ from the surface. 
This correlation corresponds to $\dir{x} \cdot \fieldcorrnor^{EE}(z_s \dir{z}, \rho \dir{x} + z_s \dir{z}) \cdot \dir{x}$ with $\dir{z}$ perpendicular to the surface and $\dir{x}$ parallel to it. The results are displayed for the extreme near field ($z_s = 0.02 \lambda$, Fig. \ref{fig:01}(a)), the near-field  ($z_s = 0.5\lambda$, Fig. \ref{fig:01}(b)) and the far-field ($z_s = 10 \lambda$, Fig. \ref{fig:01}(c)) ranges. The results obtained concerning the lateral correlation of the thermal fields in these three regimes match well those from~\cite{Lau_2007} and can be explained in a similar way. Note that the correlation function and the CDOS have been normalized separately. Due to symmetry reasons, the lateral correlation of the fields is purely real so that only its real part is displayed.

Looking at the extreme near field (Fig. \ref{fig:01}(a)), two different regimes can be noticed. The first one corresponds to the sharp peak for $\rho < z_s$. Due to the close proximity with the surface, the evanescent part of spectrum greatly enhances the intensity of the fields thermally emitted. However, this enhancement is very local and the correlation rapidly drops with distance~\cite{Lau_2007}. The second regime corresponds to $\rho > \lambda$, where slowly decaying oscillations are visible. This long-range correlation is the signature of the surface plasmon supported by the interface ~\cite{Carminati_1999}. At that distance from the interface, the CDOS looks very similar to the correlation of the thermal fields. Indeed, the CDOS can be decomposed into two contributions: one from the SiC and one from the background. Close to the interface, the background contribution, which is only consisting of the propagative part of the spectrum, is very small compared to that of the SiC.

Going farther away from the interface (Fig. \ref{fig:01}(b)), the sharp peak that existed in the close vicinity of the surface decreases and widens out as the highly evanescent part of the spectrum is progressively filtered out. At a distance of $z_s = 0.5 \lambda$, the surface plasmon constitutes the dominant contribution to the thermal fields, hence the slowly decaying oscillations. The CDOS, however, exhibits a different behaviour. The reason is that at these frequency and distance from the surface, a SiC half-space with a flat surface makes a poor absorber and thus a poor emitter. The contribution of the SiC to the thermal field is thus small compared to that of the incident blackbody radiation and its reflection on the sample.

Last, at a large distance from the surface (Fig. \ref{fig:01}(c)), the whole evanescent part of the spectrum has been filtered out, including the surface plasmon. The emissivity of the SiC half-space varying slowly with respect to the angle of emission, the shape of the correlation of the fields it emits is roughly similar to that of a black body. Thus, the CDOS, which is mainly originating from the incident blackbody radiation, exhibits a similar shape. 

{\color{black}In order to illustrate Eq. \eqref{eqbis:08} for luminescent emitters, we consider one further example consisting of a light emitting diode (LED) working at free-space wavelength of 800nm. The LED is made of a GaAs/Al$_{0.3}$Ga$_{0.7}$As heterostructure, as illustrated in the inset of Fig. \ref{fig:04}. The active zone corresponds to a 10nm thick GaAs layer. The GaAs layer is sandwiched between two layers of Al$_{0.3}$Ga$_{0.7}$As and acts as a quantum well \cite{Rosencher_2002}. On top of the structure, a 25nm thick silver layer is deposited. Since silver exhibits a plasmonic behaviour at the frequency considered, it can support two different surface modes \cite{Sarid_1981, Inagaki_1985}, which correspond to the hybridized version of the surface plasmons supported by each interface. Thus, one of the two surface modes can leak into the semiconductor, while the other one is evanescent in both the air and the semiconductor. Combined with a proper patterning, this kind of structure has been used to shape the radiation pattern of the LED \cite{Kock_1990}. 

At the frequency considered, the relative permittivities of silver and Al$_{0.3}$Ga$_{0.7}$As are $\varepsilon_r = -27.2+0.98j$ \cite{Johnson_1972} and $\varepsilon_r = 11.75$ \cite{Aspnes_1986}, respectively. The absorption of a normally incident plane wave by the quantum well is about 0.6\% \cite{Rosencher_2002}. Losses in the quantum well are modeled using the complex permittivity $\varepsilon_r = 11.75 + 0.25j$, whose imaginary part has been chosen to match this result. 

We study the excitation of the surface modes by the active zone when it is located 90nm and 200nm below the silver layer. To do so, we consider the $\dir{z}\dir{z}$ component of the mixed losses of the quantum well for sources located 10nm above the silver layer. The results are displayed in Fig. \ref{fig:04}. When the quantum well is located 90nm below the silver layer, losses in the quantum well are mediated by both surface modes. Thus, Eq. \eqref{eqbis:08} tells us that the quantum well can efficiently excite both surface modes. However, if the quantum well is located 200nm away from the silver layer, the mixed losses are smoother, due to the poor coupling between the most evanescent surface mode and the quantum well. Using Eq. \eqref{eqbis:08}, one can deduce that the quantum well cannot excite efficiently the most evanescent surface mode at such a distance.

\begin{figure}
\center
\includegraphics[width = 8cm]{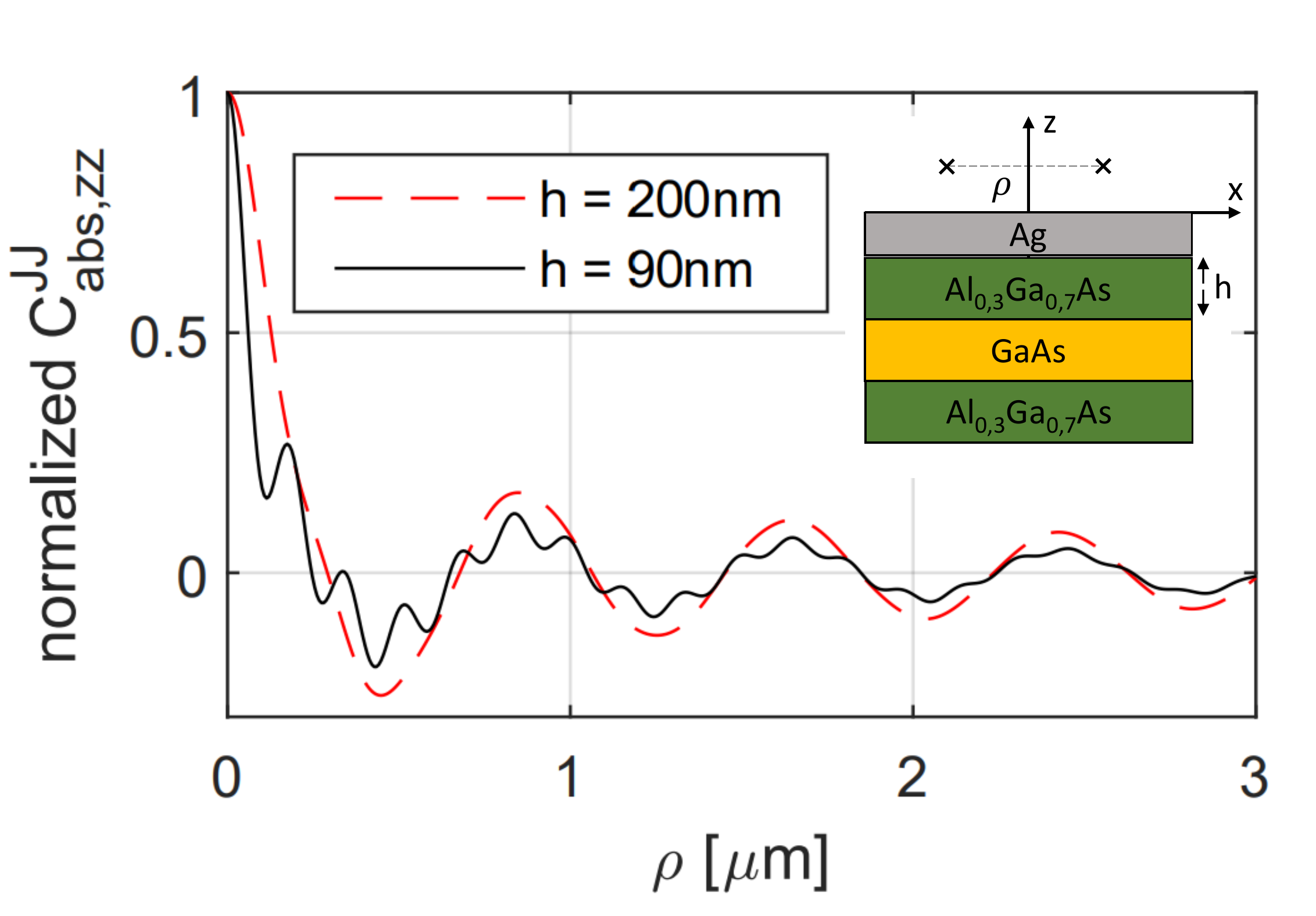}
\caption{Normalized two-points correlation of the fields emitted by the active zone of the LED, when the active zone is located 90nm and 200nm below the metal contact (black solid line and red dashed line, respectively). }
\label{fig:04}
\end{figure} 
}

\section*{Conclusion}
To conclude, we have generalized a coherence-absorption relation to non-equilibrium systems. Provided that the temperature and potential associated to excitation of a coupling mechanism are uniform over a given region, a simple reciprocity relation exists between the correlation of the fields emitted from this region by this coupling mechanism and the associated mixed losses. This generalized coherence-absorption relation gives access to the coherence properties of light emitted by sources by simply computing or measuring losses. Thus, it provides a natural framework to explore coherence properties of light emitted by different sources (atoms, molecules, quantum dots, semiconductors, hot electrons) embedded in complex environments such as metasurfaces, antennas, cavities, waveguides or disordered systems. Finally, we note that the coherence absorption relation has been introduced in the context of optics. It could be implemented in radiowaves by extending the techniques developed to study the local density of states in metasurfaces \cite{Rustomji_2017}.

\section*{Acknowledgement}
We thank Christophe Craeye for his comments and for the fruitful discussion we had.

\section*{Fundings}
D. Tihon thanks the European Union's Horizon 2020 research and innovation programme for funding him under the Marie Sk\l odowska-Curie grant agreement No. 842184.
J.-J. Greffet, E. Bailly and B. Vest thank Agence Nationale de la Recherche for the Grant ANR-17-CE24-0046.

\section*{Disclosure}
The authors declare no conflicts of interest.

%%%%%%%%%%%%%%%%% Supplementaries %%%%%%%%%%%%%%%%%%%%%%
\section*{Appendix}
\subsection*{A. Power absorption vs. spontaneous emission}
Since the proposed method is a spectral method, an $\exp(-j\omega t)$ time dependency of the fields and currents is assumed. The frequency dependence of the different quantities hereafter is implicit.

In the following, we consider a lossy structure described using position-dependent complex permittivity $\varepsilon(\rpos) = \varepsilon'(\rpos) + j \varepsilon"(\rpos)$ and real permeability $\mu(\rpos)$. As explained in the body of the paper, non-local effect are not considered in this paper. For the sake of clarity, only electric losses are considered. However, the extension to magnetic losses is straightforward.

If the structure is excited using current sources (magnetic currents $\vect{M}(\rpos)$ or electric currents $\vect{J}(\rpos)$), electric ($\vect{E}(\rpos)$) and magnetic ($\vect{H}(\rpos)$) fields will be generated. We introduce the Green's tensor $\GF(\rpos, \rposp)$ of the structure such that
\begin{align}
\label{eq:61}
\vect{E}(\rpos) &= \iiint \Big[ 
\GF^{EJ}(\rpos, \rposp) \cdot \vect{J}(\rposp) 
+
\GF^{EM}(\rpos, \rposp) \cdot \vect{M}(\rposp) 
\Big] dV(\rposp)
\\
\label{eq:62}
\vect{H}(\rpos) &= \iiint \Big[ 
\GF^{HJ}(\rpos, \rposp) \cdot \vect{J}(\rposp) 
+
\GF^{HM}(\rpos, \rposp) \cdot \vect{M}(\rposp) 
\Big] dV(\rposp)
\end{align}
where the integral is carried over the whole space.

One key tool to establish the relation between the cross-spectral density tensor and the mixed losses associated to the structure is the use of Lorentz reciprocity \cite{Collin_1960}. As explained in \cite{Rodriguez_Aug2013}, this reciprocity theorem can be extended to magnetic currents and fields to provide:

\begin{equation}
\label{eq:01}
\begin{soe}
\GF^{EJ}(\rone, \rtwo) = \Big(\GF^{EJ} \Big)^T(\rtwo, \rone), \\
\GF^{EM}(\rone, \rtwo) = -\Big(\GF^{HJ} \Big)^T(\rtwo, \rone), \\
\GF^{HJ}(\rone, \rtwo) = -\Big(\GF^{EM} \Big)^T(\rtwo, \rone), \\
\GF^{HM}(\rone, \rtwo) = \Big(\GF^{HM} \Big)^T(\rtwo, \rone),
\end{soe}
\end{equation}
with $\mat{A}^T$ corresponding to the transpose of dyadic $\mat{A}$.

Using this reciprocity relation and the fluctuation-dissipation theorem, it is possible to find a direct relation between the power absorbed through a given coupling mechanism in a structure and the correlation of the fields emitted through the excitation of this coupling mechanism. In order to bring out this relation, we will first study separately the correlation of the fields emitted by a chosen coupling mechanism and then the mixed losses associated to that same coupling mechanism. Then, we will use the Lorentz reciprocity relation to provide the link between both. 

\subsubsection*{A.1 Spontaneous emission}
Due to the finite temperature, random fluctuations of current will appear in the volume and radiate fields. The goal of this Section is to characterize the spatial correlation of these fields averaged over time. If the temperature potentials and material parameters do not vary with time, the current fluctuations are stationary and the time average can be replaced with an ensemble average (see \cite{Rytov_1989,Born_1991,Greffet_Apr2018}). The time averaged complex correlation of the fields thus corresponds to the cross-spectral density tensor of the fields $\fieldcorr$ \cite{Greffet_Apr2018} given by 
\begin{align}
\label{eq:02}
\fieldcorr 
&= 
\left\langle 
	\matrix{\vect{E}\\ \vect{H}}(\rone) 
	\cdot 
	\matrix{\vect{E} \\ \vect{H}}^\dagger(\rtwo)
	\right\rangle, 
\end{align}
with the superscript $\dagger$ corresponding to the transpose conjugate operation. 

These electric and magnetic fields are generated by randomly induced current sources inside the lossy volume. For a given realization of the ensemble used in \eqref{eq:02}, the field generated by these fluctuations can be obtained using the Green's tensor. Integrating the contribution of all the random sources in the volume gives
\begin{equation}
\begin{split}
\fieldcorr &=
\Bigg\langle 
	\Big[ \iiint_\Omega \GF^J(\rone, \rthree) \cdot \vect{J}(\rthree) dV(\rthree) \Big]  \cdot \Big[\iiint_\Omega \GF^J(\rtwo, \rfour) \cdot \vect{J}(\rfour) dV(\rfour) \Big]^\dagger
\Bigg\rangle,
\end{split}
\end{equation}
with $\Omega$ the volume occupied by the emitter and 
\begin{equation}
\GF^J(\rone, \rtwo)
=
\matrix{\GF^{EJ}\\
\GF^{HJ}}(\rone, \rtwo).
\end{equation}

The convolution of a current distribution by the Green's tensor is a linear operation, so that the ensemble average and the convolution operations order can be swapped:
\begin{equation}
\label{eq:04}
\begin{split}
\fieldcorr & = \iiint_\Omega \iiint_\Omega
\GF^J(\rone, \rthree) 
\cdot 
\Big\langle \vect{J}(\rthree) \cdot \vect{J}^\dagger(\rfour) \Big\rangle
\cdot 
\big(\GF^J\big)^\dagger(\rtwo, \rfour)
~ dV(\rthree) ~ dV(\rfour).
\end{split}
\end{equation}

The great advantage of \eqref{eq:04} is that the correlation of the fields is expressed as a function of the correlation of the fluctuating currents, which can be expressed using the fluctuation-dissipation theorem \cite{Landau_1969,Greffet_Apr2018}. For an homogeneous medium, it leads to
\begin{equation}
\label{eq:05}
\Big\langle
\vect{J}(\rthree) \cdot \vect{J}^\dagger(\rfour) 
\Big\rangle
=
2 \omega \sum_q \varepsilon_q"(\rthree) \, \Theta(T_q(\rthree), \mu_q(\rthree)) \, \delta(\rthree-\rfour) \, \Imat,
\end{equation}
with $\Imat$ the 3x3 identity matrix, $\delta$ the Dirac delta function and $\Theta(T_q, \mu_q)$ that is given by
\begin{equation}
\label{eq:73}
\Theta(T_q, \mu_q) = \dfrac{\hbar \omega}{\exp\bigg(\dfrac{\hbar \omega-\mu_q}{k_B T_q}\bigg) -1}.
\end{equation}

Substituting \eqref{eq:05} into \eqref{eq:04}, switching the integration and summation orders and simplifying the integrals gives
\begin{align}
\label{eq:51}
\fieldcorr & = \sum_q
2 \omega
 \iiint_\Omega  \Theta\big(T_q(\rthree), \mu_q(\rthree)\big)  \varepsilon_q"(\rthree)
\GF^J(\rone, \rthree) \cdot \big(\GF^J\big)^\dagger(\rtwo, \rthree)
dV(\rthree) \\
& \triangleq \sum_q \mat{C}_{\text{em},q}(\rone, \rtwo) \\
& \triangleq \sum_q 
\matrix{
\mat{C}_{\text{em},q}^{EE} \hspace{0.3cm} 
\mat{C}_{\text{em},q}^{EH}  \\
~ \vspace{-0.5cm} \\ 
\mat{C}_{\text{em},q}^{HE} \hspace{0.3cm} \mat{C}^{HH}_{\text{em},q}
}(\rone, \rtwo).
\end{align}

To summarize, $\mat{C}_{\text{em},q}^{ij}(\rone, \rtwo)$ provides the complex correlation between the fields of type $i$ generated by coupling mechanism $q$ in $\rone$ and fields of type $j$ generated by the same coupling mechanism $q$ in $\rtwo$. 

If the temperature and chemical potential associated to the excitation of the coupling mechanism are uniform on the emissive zone $\Omega$, one obtains
\begin{equation}
\label{eq:emission}
\mat{C}_{\text{em},q}^{ij}(\rone, \rtwo) 
=
2\omega \Theta(T_q,\mu_q)  \iiint_\Omega \varepsilon_q"(\rthree) \mat{\Gamma}^{iJ}(\rone, \rthree) \cdot \Big(\mat{\Gamma}^{jJ}\Big)^\dagger(\rtwo, \rthree) dV(\rthree).
\end{equation}

Note that we have introduced the concept of coupling mechanism associated to a contribution to the imaginary part of the permittivity:
it corresponds to the absorption of a photon that is triggered by its interaction with a set $q$ of one or many particles (electron, hole, phonon). 

Here we stress a difference between local thermodynamic quasi-equilibrium and non-equilibrium situations regarding the coupling mechanism. We take as an example the absorption of a photon by a semiconductor. 
The set $q$ of particles involved in an interband transition consists of one electron in the valence band, one hole in the conduction band and possibly one or many absorbed or generated phonons in case of indirect transition. At equilibrium, the distribution of electrons (and thus holes) in the semiconductor is given by a Fermi-Dirac distribution with well-defined temperature $T_q$ and chemical potential. Similarly, the energy distribution of phonons is given by a Bose-Einstein distribution with the same temperature $T_q$. 

When dealing with quasi-equilibrium situations,we may have to introduce different chemical potentials for electrons in the conduction band and in the valence band (the so-called quasi Fermi levels) resulting in the introduction of a photon chemical potential $\mu_q$ in Equation \eqref{eq:73}. We also point out that it may be necessary to define different temperatures for particles involved (electrons of the two bands or phonons). Then, the ratio between emission and absorption becomes dependent on the fine structure of the absorbing material (density of states of each particle involved). Thus Equation \eqref{eq:05} is not valid anymore and needs to be modified as discussed in \cite{Guillemoles_2016}.

\subsubsection*{A.2 Power absorption}
Now, we consider the power absorbed by the structure studied previously when it is excited by two different current sources. The absorptive behaviour of the structure can be described using its mixed losses tensor $\Cabsnor$ \cite{Rytov_1989}. The development used here is similar to the one presented in \cite{Craeye_2014, Tihon_2016, Withington_2017}.

The structure is illuminated by electric and magnetic current sources corresponding to a current distribution of $[\vect{J}^T, \vect{M}^T]^T(\rpos)$. The electric field generated by these sources at any position $\rthree$ can be computed using the Green's tensor defined in \eqref{eq:61} and \eqref{eq:62}:
\begin{equation}
\label{eq:07}
\vect{E} (\rthree)
= 
\iiint
\GF^E(\rthree, \rone) 
\cdot 
\vectcurr(\rone)
dV(\rone)
\end{equation}
with
\begin{equation}
\GF^E(\rthree, \rone)
=
\matrix{ \GFej \hspace{0.3cm} \GFem}(\rthree, \rone).
\end{equation}

The total power $\Pabsq$ absorbed by the structure through coupling mechanism $q$ corresponds to the integral over the structure of the power absorbed locally
\begin{equation}
\label{eq:08}
\Pabsq = \dfrac{\omega}{2} \iiint_\Omega \varepsilon_q"(\rthree) \vect{E}^\dagger(\rthree) \cdot \vect{E}(\rthree) dV(\rthree).
\end{equation}
Note that, as mentioned earlier, no magnetic losses and no non-local phenomena is considered.
Substituting \eqref{eq:07} into \eqref{eq:08} leads to
\begin{equation}
\begin{split}
\Pabsq = \dfrac{\omega}{2} 
\iiint_\Omega \varepsilon_q"(\rthree) 
&
\Bigg[
\iiint \GF^E(\rthree, \rone) \cdot 
\vectcurr
(\rone) dV(\rone) \Bigg]^\dagger \\
&
\cdot \Bigg[ \iiint \GF^E(\rthree, \rtwo) \cdot \vectcurr(\rtwo) dV(\rtwo) \Bigg] dV(\rthree)
\end{split}
\end{equation}

Reordering the integration order and distributing the transpose-conjugate operator leads to 
\begin{equation}
\label{eq:71}
\Pabsq = \iiint \iiint \vectcurr^\dagger (\rone) \cdot \Cabsq \cdot \vectcurr(\rtwo) \, dV(\rone) \, dV(\rtwo)
\end{equation}
with
\begin{equation}
\label{eq:52}
\Cabsq = \dfrac{\omega}{2} \iiint_\Omega \varepsilon_q"(\rthree) \GF^{E,\dagger}(\rthree, \rone) \cdot \GF^E(\rthree, \rtwo) \, dV(\rthree).
\end{equation}

In the particular case where the sources exciting the structure consist of two localized current sources $[\vect{J}_1^T, \vect{M}_1^T]^T$ and $[\vect{J}_2^T, \vect{M}_2^T]^T$ located in position $\rone$ and $\rtwo$, respectively, the total current distribution reads
\begin{equation}
\vectcurr(\rpos) = 
\vectcurrone \delta(\rpos-\rone) 
+
\vectcurrtwo \delta(\rpos-\rtwo)
\end{equation}
and Equation \eqref{eq:71} becomes
\begin{equation}
\label{eq:72}
\Pabsq = 
\sum_{i,j = 1}^{2}
\matrix{\vect{J}_i \\ \vect{M}_i}^\dagger
\cdot 
\Cabsqnor(\rpos_i, \rpos_j)
\cdot
\matrix{\vect{J}_j \\ \vect{M}_j}.
\end{equation}

Finally, the cross-spectral power density tensor can be decomposed into four different parts corresponding to four possible combinations of electric and magnetic current pairs:
\begin{equation}
\label{eq:50}
\Cabsqnor(\rpos_i, \rpos_j)
=
\matrix{\Cabsqnor^{JJ} \hspace{0.3cm} \Cabsqnor^{JM} \\ ~ \\ \Cabsqnor^{MJ} \hspace{0.3cm} \Cabsqnor^{MM}}(\rpos_i, \rpos_j).
\end{equation}
Isolating the sole contribution of the electric currents in \eqref{eq:72} leads to Equation (5).

To summarize this result, $\Cabsqnor(\rone, \rtwo)$ corresponds to the mixed losses of the structure for coupling mechanism $q$ when it is illuminated by electric and/or magnetic current sources. More specifically, entry $(k,k)$ of $\Cabsqnor^{ii}(\rone, \rone)$ corresponds to the power absorbed by the structure through coupling mechanism $q$ when it is illuminated by a monochromatic dipole-like impressed electric ($i=J$) or magnetic ($i=M$) current source oriented along direction $k$ and located at position $\rone$. Entry $(k,l)$ of $\Cabsqnor(\rone, \rtwo)$ corresponds to the cross-term in the total power absorbed by the structure, which is related to interferences, when it is illuminated by a $k$-directed current of type $i$ located at position $\rone$ and a $l$-directed current of type $j$ located at position $\rtwo$, $i$ and $j$ corresponding to electric and/or magnetic currents and $k$ and $l$ corresponding to any of the three directions of space (\eg \xdir, \ydir and \zdir directions using a Cartesian coordinate system).

\subsubsection*{A.3 Extended reciprocity}
It can be noticed that Equations \eqref{eq:emission} and \eqref{eq:52} look very similar. Using Lorentz reciprocity, it is possible to express one as a function of the other. For simplicity, we only treat the $\fieldcorrqEH$ and $\Cabsqjm$ terms, but other terms can be treated similarly.

Substituting \eqref{eq:01} into \eqref{eq:emission}, one obtains
\begin{subequations}
\begin{align}
\fieldcorrqEH 
& =
2\omega \Theta(T_q, \mu_q) 
\iiint_\Omega \varepsilon_q"(\rthree) \GFej(\rone, \rthree) \cdot \big(\GFhj\big)^\dagger(\rtwo, \rthree) dV(\rthree)
\\
& =
2\omega \Theta(T_q, \mu_q) 
\iiint_\Omega \varepsilon_q"(\rthree) \big(\GFej\big)^{T}(\rthree, \rone) \cdot \big(-\GFem\big)^*(\rthree, \rtwo) dV(\rthree)
\\
& =
-4 \Theta(T_q, \mu_q) \Bigg(\dfrac{\omega}{2}
\iiint_\Omega \varepsilon_q"(\rthree) \big(\GFej\big)^{\dagger}(\rthree, \rone) \cdot \big(\GFem\big)(\rthree, \rtwo) dV(\rthree) \Bigg)^*
\\
& =
-4 \Theta(T_q, \mu_q) \Cabsqjmconj
\end{align}
\end{subequations}
with $\mat{A}^*$ the element-wise complex conjugate of $\mat{A}$. Applying similar treatment to the other terms of \eqref{eq:51}, one finds that
\begin{equation}
\label{eq:10}
\begin{soe}
\fieldcorrqEE = 4 \Theta(T_q, \mu_q) \Cabsqjjconj, \\ ~ \vspace{-0.2cm} \\
\fieldcorrqEH = - 4 \Theta(T_q, \mu_q) \Cabsqjmconj, \\ ~ \vspace{-0.2cm} \\
\fieldcorrqHE = - 4 \Theta(T_q, \mu_q) \Cabsqmjconj, \\ ~ \vspace{-0.2cm} \\
\fieldcorrqHH = 4 \Theta(T_q, \mu_q) \Cabsqmmconj. 
\end{soe}
\end{equation}

Equation \eqref{eq:10} provides the relation between the power absorbed through loss mechanism $q$ by a region of space $\Omega$ when it is illuminated by electric or magnetic currents sources and the correlation of the electric and magnetic fields spontaneously emitted by this same region of space by loss mechanism $q$ when its temperature and chemical potential are uniform over $\Omega$.

\subsection*{B. Alternative derivation of the CDOS}
The CDOS \cite{Caze_2013}, which describes the fields correlation when the whole space is at thermodynamic equilibrium, can be obtained using the fluctuation-dissipation theorem in the same way as the correlation of the current density thermally generated was obtained. Alternatively, it is possible to retrieve the CDOS using the reciprocity relations of \eqref{eq:10}. Consider two electric current sources $\vect{J}_1$ and $\vect{J}_2$ located at positions $\rone$ and $\rtwo$, respectively. Since we are interested in the fields generated by the whole space at thermal equilibrium, we look at the power dissipated by the whole space, including radiation losses. The total power dissipated corresponds to the power generated by these two current sources:
\begin{equation}
P_\text{tot} = -\dfrac{1}{2} \text{Re}\{\vect{J}_1^\dagger \cdot \vect{E}(\rone) + \vect{J}_2^\dagger \cdot \vect{E}(\rtwo) \}
\end{equation}
Using the Green's tensor defined in \eqref{eq:01} one obtains
\begin{align}
P_\text{tot} = -\dfrac{1}{2}  &\text{Re}\big\{\vect{J}_1^\dagger \cdot (\GFej(\rone, \rone) \cdot \vect{J}_1 + \GFej(\rone, \rtwo) \cdot \vect{J}_2) 
\nonumber \\
& \hspace{0.2cm}
 + \vect{J}_2^\dagger \cdot (\GFej(\rtwo, \rone) \cdot \vect{J}_1 + \GFej(\rtwo, \rtwo) \cdot \vect{J}_2) \big\}.
\end{align}
Using the identity $2\text{Re}\{A\} = A + A^*$ gives
\begin{equation}
\label{eq:11}
\begin{split}
P_\text{tot} = -\sum_{i,j=1,2} \dfrac{1}{4} \Big( 
& \vect{J}_i^\dagger \cdot \GFej(\rpos_i, \rpos_j) \cdot \vect{J}_j + \vect{J}_i^T \cdot \GFej^*(\rpos_i, \rpos_j) \cdot \vect{J}_j^* \Big).
\end{split}
\end{equation}
It can be noticed that the second term of the RHS of \eqref{eq:11} is a scalar, so that it is equal to its transpose. Applying the transposition, it is possible to group differently the terms of the sum, leading to
\begin{equation}
P_\text{tot} = -\sum_{i,j=1,2} \dfrac{1}{4} \Big(\vect{J}_i^\dagger \cdot \big(\GFej(\rpos_i, \rpos_j) + \GFej^\dagger(\rpos_j, \rpos_i)\big) \cdot \vect{J}_j \Big)
\end{equation}
Last, using the Lorentz reciprocity of \eqref{eq:01} and considering that $A + A^* = 2\text{Re}\{A\}$ gives
\begin{equation}
P_\text{tot} = -\dfrac{1}{2} \sum_{i,j=1,2} \vect{J}_i^\dagger \cdot \text{Re}\{\GFej(\rpos_i, \rpos_j)\} \cdot \vect{J}_j.
\end{equation}
The last result illustrates that the cross-spectral power density tensor $\Cabsnor(\rpos_i, \rpos_j)$ of the whole space corresponds to the real part of the Green's tensor, within a $1/2$ factor. Thus, using the reciprocity relation of \eqref{eq:10}, one obtains that the cross-spectral electric field density tensor of the whole space at thermodynamic equilibrium reads:
\begin{equation}
\label{eq:12}
\fieldcorrEE = -2\Theta(T_0) \text{Re}\{\GFej(\rone, \rtwo)\},
\end{equation}
with $T_0$ the uniform temperature of the space.

A similar derivation can be done for $\fieldcorrEH$, $\fieldcorrHE$ and $\fieldcorrHH$, leading to
\begin{equation}
\begin{soe}
\fieldcorrEE = -2 \Theta(T_0) \text{Re}\{\GF^{EJ}(\rone, \rtwo)\}, \\
\fieldcorrEH = -2j \Theta(T_0) \text{Im}\{\GF^{EM}(\rone, \rtwo)\}, \\
\fieldcorrHE = -2j \Theta(T_0) \text{Im}\{\GF^{HJ}(\rone, \rtwo)\}, \\
\fieldcorrHH = -2 \Theta(T_0) \text{Re}\{\GF^{HM}(\rone, \rtwo)\}.
\end{soe}
\end{equation}
with $\text{Im}\{a+jb\} = b$ for $a,b \in \mathbb{R}$.

%%%%%%%%%% If using BibTeX:
\bibliography{biblio_OptEx}

\end{document}